\documentclass[10pt,a4paper]{article}

\usepackage[utf8]{inputenc}
\usepackage{amsmath}
\usepackage{amsfonts, dsfont}
\usepackage{amsthm}
\usepackage{amssymb}
\usepackage{graphicx}
\usepackage{natbib}
\usepackage{booktabs}
\usepackage[margin=1in]{geometry}
\usepackage{float}
\usepackage[flushleft]{threeparttable}
\usepackage[colorlinks=true, allcolors=blue]{hyperref}
\usepackage{tikz}
\usetikzlibrary{arrows, automata}

\usepackage[ruled]{algorithm2e}

\newtheorem{prop}{Proposition}

\newtheorem{assumption}{Assumption}

\newtheorem*{implication}{Implication}

\newcommand{\cI}{\mathcal{I}}
\newcommand{\cA}{\mathcal{A}}
\newcommand{\cP}{\mathcal{P}}
\newcommand{\cS}{\mathcal{S}}

\newcommand{\Normal}{\operatorname{N}}
\DeclareMathOperator{\var}{Var}
\DeclareMathOperator{\cov}{Cov}
\newcommand{\LATE}{\mathsf{LATE}}
\newcommand{\ITThat}{\operatorname{\widehat{ITT}}}
\newcommand{\ITT}{\operatorname{ITT}}

\title{Auction Throttling and Causal Inference of Online Advertising Effects}
\author{
George Gui\thanks{Graduate School of Business, Stanford University, \href{mailto:ggui@stanford.edu}{ggui@stanford.edu}}, Harikesh Nair\thanks{Graduate School of Business, Stanford University, \href{mailto:hnair@stanford.edu}{hnair@stanford.edu}}, Fengshi Niu \thanks{Graduate School of Business, Stanford University, \href{mailto:fniu@stanford.edu}{fniu@stanford.edu}\newline Acknowledgement: We thank Ron Berman, Carlos Carrion, Han Hong, Guido Imbens, Yewon Kim, Jack Lin, Xiliang Lin, Navdeep Sahni, Di Wu, Paul Yan, and seminar participants at Stanford Marketing WIP, Stanford Econometrics Lunch, and Marketing Science Conference 2021 for their generous support and insightful comments. All errors are our own.}
}
\date{First draft: Jan 3, 2022 \\ This draft: Feb 16, 2022}

\begin{document}
\maketitle

\noindent \textbf{Abstract:}  
Causally identifying the effect of digital advertising is challenging, because experimentation is expensive, and observational data lacks random variation. This paper identifies a pervasive source of naturally occurring, quasi-experimental variation in user-level ad-exposure in digital advertising campaigns. It shows how this variation can be utilized by ad-publishers to identify the causal effect of advertising campaigns. The variation pertains to auction throttling, a probabilistic method of budget pacing that is widely used to spread an ad-campaign’s budget over its deployed duration, so that the campaign’s budget is not exceeded or overly concentrated in any one period. The throttling mechanism is implemented by computing a participation probability based on the campaign’s budget spending rate and then including the campaign in a random subset of available ad-auctions each period according to this probability. We show that access to logged-participation probabilities enables identifying the local average treatment effect (LATE) in the ad-campaign. We present a new estimator that leverages this identification strategy and outline a bootstrap procedure for quantifying its variability. We apply our method to real-world ad-campaign data from an e-commerce advertising platform, which uses such throttling for budget pacing. We show our estimate is statistically different from estimates derived using other standard observational methods such as OLS and two-stage least squares estimators. Our estimated conversion lift is 110\%, a more plausible number than 600\%, the conversion lifts estimated using naive observational methods.

\section{Introduction}

Measuring the causal impact of online advertising is important for both advertisers and digital advertising platforms. For advertisers, understanding the impact of advertising is essential for evaluating past campaign performance and optimizing future spending. Comparing the causal effect to the cost of the campaign helps the advertiser assess the incremental return on investment from the campaign, and is a critical determinant of whether or not the campaign should be continued (e.g. \cite{blake_consumer_2015} for a well known example with \texttt{eBay}). The causal effects are also important determinants of how much advertisers should bid in real-time bid (RTB) auctions, because they represent the valuation of the advertiser for the average impression acquired in the campaign. As such, the ability to measure incremental ad-effects is now considered a key driver of the success of a firm's digital advertising strategy (e.g., \cite{gordon_comparison_2019}). For platforms, causally measuring ad-effects is important for valuing their own advertising inventory, for improving their ad-selling mechanisms, and for creating automation solutions that help advertisers do better bidding and budgeting of their campaigns on the platform (e.g., \citet{borgs2007dynamics,chakrabarty2007budget,feldman2007budget,muthukrishnan2007stochastic,muthukrishnan2010stochastic,amin2012budget,zhang_joint_2012,nuara2018combinatorial,geng_automated_2021}). 

However, measuring the causal impact of digital advertising is challenging. Typical observational ad data lack the variation required to measure causal effects because ad exposure is not randomly assigned, and significant biases remain even after matching exposed and unexposed users on observable dimensions (see \cite{gordon_comparison_2019}). Although running experiments seem like a viable alternative, it is often expensive. It is well known that the causal effects of digital ads are small in magnitude, and large sample sizes are required to measure it with statistical precision (e.g., \cite{Sahni2015,lewis_unfavorable_2015}. Large-scale experimentation can be economically unfavorable for advertisers because under randomization, ads are withheld from users in the control group, some of whom would have responded favorably to the ads had they been targeted. Large scale randomization may also hurt ad platforms because it requires their users to see random ads that may not match their preferences. The problem is more challenging if advertisers are interested in optimizing within-campaign expenditure: this requires determining whether to spend more on certain subpopulations and less on other subpopulations. Solving this requires measuring heterogeneous treatment effects of advertising, which requires more data than measuring average effects. Given the high costs of experimentation, it is valuable to identify alternative source of variation in observational digital advertising data that can help pin down causal ad-effects. This is the goal of this paper. 

We identify a pervasive source of naturally occurring, quasi-experimental variation in user-level ad-exposure in digital advertising campaigns. We show how this variation can be utilized by ad-platforms and ad-publishers to identify the causal effect of advertising campaigns. Our proposal leverages an ad-serving strategy referred to as ``auction throttling'' which is widely used by major ad publishers such as Google (\cite{karande_optimizing_2013} and \cite{balseiro_budget_2017}), Linkedin (\cite{agarwal_budget_2014}) and others, and which has been well studied theoretically in the literature (e.g., \cite{charles2013budget}). 

Auction throttling is a budget smoothing mechanism motivated by the fact that advertising campaigns typically have limited budgets. The budget limitation prevents advertisers from participating in all eligible auctions, i.e. auctions for impressions that match the campaign's targeting criteria. The concern is that if advertisers participate in every eligible auction early in the campaign, they will exhaust their budget too early and lose the opportunity to target customers that arrive later in  the campaign. Hence, it may be in the advertiser's interest to smooth their budget over the campaign duration. This smoothing process is referred to as ``pacing.'' Pacing budgets is also helpful for the ad-platform because in its absence, auction pressure on the platform becomes highly variable. For instance, if most ad campaign budgets are spent early in the day, auction pressure (and consequently ad-prices) become high early in the day and low later in the day. Understanding this, advertisers may shift budgets to later in the day leading to complex spending and pricing variability and auction dynamics on the platform. Pacing the budget helps the platform manage this variability.

The throttling mechanism paces the spending for a given advertising campaign, by calculating a participation probability for the campaign for each auctioned impression. This probability is calculated by the platform's ad-server using real-time bidding information, such as the recent spending rate, remaining budget, remaining campaign time etc. The specifics of the implementation vary, but broadly speaking, the participation probability is high when the remaining budget is high or recent spending rate is low, and is low when the remaining budget is low or recent spending rates are high. Throttling is achieved by including campaigns \textit{randomly} into auctions with this participation probability. Conditional on this, auctions proceed as usual from that point onwards with the campaign's bids included (when its throttled in), or without the campaign's bids (when its throttled out). Dynamically adjusting the participation probability over time helps pace the campaign. The participation probabilities are usually logged by platforms.\footnote{An alternative mechanism to pace the budget is to lower the bid when the budget is low. \cite{balseiro_budget_2017} has compared the revenue implications of these different mechanisms. Our paper can be understood as demonstrating an extra benefit of probabilistic throttling: providing a source of pervasive identifying variation in the data that does not require running experiments.}

Our interest is in leveraging the participation probabilities for causal inference. We emphasize our interest here is not in analyzing the pacing algorithm or its design or implications. We show that typical implementations of such pacing  mechanisms serve as a source of naturally occurring, quasi-experimental variation in auction participation, which induces random variation in user-level ad exposure. Typically, auction participation cannot  be regarded as exogenous to post-auction outcomes because it depends on many factors that correlate with outcomes, such as ad targeting criteria, as well as remaining campaign budgets (which is affected by the arrival of past users). Nevertheless, we show that auction participation is \textit{conditionally} independent of the potential outcomes of interests after conditioning on this participation probability. The implication is that ad-platforms and publishers can use the participation induced by the throttling mechanism as a \textit{conditional instrument variable} for identifying the impact of advertising campaign, as well as its heterogeneous treatment effects over different sub-populations. 

We articulate the key requirements of a throttling algorithm to facilitate such inference, and make precise the specific causal estimand that can be estimated leveraging this variation. 
We present an estimator that learns this estimand from typical throttled ad-campaign data. 
We propose an analytical standard error that provably works under a fixed throttling regime and a bootstrap standard error that works generally when we know the throttling algorithm. 
We use both simulation and empirical data from an e-commerce advertising platform to demonstrate the procedure and to benchmark it against alternative approaches. 
To the best of our knowledge, the value of this variation for inference has not been articulated in the literature, and therefore, we believe our proposal is new.

Our procedure requires access to the logged participation probabilities and knowledge of the throttling rule. Therefore, it should be seen as a way to implement causal inference of RTB ads from the publisher/platform's perspective. For platforms and publishers who implement budget pacing methods already via probabilistic throttling, our procedure provides a way to automatically deliver inference on the ad campaigns on the platforms without the need for additional experimentation. Due to this reason, we believe the method presented here has practical value and  appeal.  


The remainder of the paper is organized as follows: Section 2 reviews the literature. Section 3 presents the setup and key assumptions related to the throttling mechanism. Section 4 describes the estimation procedure that leverages data generated under the throttling mechanism. Section 5 demonstrates the value of the data and the corresponding estimation procedure using simulation. Section 6 discusses the inference and standard errors. Section 7 applies the method to a dataset from an e-commerce advertising platform and demonstrate the usefulness of this variation. Section 8 concludes.

\section{Literature Review}

This paper contributes to the literature on measuring the effects of online advertising. Our paper contributes to this literature by identifying a pervasive source of quasi-experimental variation in observational data that allows ad publishers to measure advertising effects without running large-scale experiments. The quasi-experimental variation is induced by the limited budget and the probabilistic throttling algorithm that is widely used by many ad publishers. 

There is now a large stream of literature (e.g., see \cite{gordon2021inefficiencies} and papers cited there) that has proposed experimental approaches for causal measurement of digital ad effects. Several of these experimental designs are from the advertisers' perspective: e.g., geo-level randomization (e.g., \cite{blake_consumer_2015}) or inducing randomization of ad-intensities by manipulating ad campaign frequency caps on DSPs (e.g., \cite{sahni2019retargeted}). Unlike these papers, our design does not require additional randomization, leverages existing ad-serving mechanisms (i.e., throttling), and is developed from the publisher/platform's perspective, rather than the advertiser's.  Our method is related to \cite{johnson_ghost_2017} which enables ad publishers to more efficiently measure advertising effect by counterfactual ad logging.  Similar to their setting, our paper focus on measuring advertising effects for ad publishers who often have direct access to the algorithm that generates advertising exposure. In contrast, we do not require additional experiments, because the probabilistic nature of the throttling algorithm generates random variation.\footnote{\cite{shapiro_positive_2018,shapiro2021tv} and \cite{stephens2017super,hartmann2018super} also leverage quasi-experimental variation to measure the effects of advertising. To assess the impact of TV advertising, the former two papers exploit the discontinuity induced in advertising along the borders of TV markets; and the latter two, exploit the variation induced in TV viewership due to changes in the identity of SuperBowl teams. In contrast, we focus on the variation induced by limited budget and the randomization induced by pacing algorithms. Our method can be applied in RTB digital advertising, while the above methods can be mostly applied to TV advertising.}  

Our estimation method is closely related to the literature on instrumental variables (IV) and the estimation of local average treatment effects (LATE). As in the IV model in \cite{ImbensAngrist1994, AngristImbensRubin1996} and the local IV model in \cite{Abadie2003, Frolich2007, Hong2010}, identification comes from a binary instrument, i.e. auction participation, that induces exogenous selection into treatment, i.e. ad exposure, for some subset of the population. However, one difference is the IV identification argument works in our setting only conditionally given the participation probability, which motivates different estimators. Our weighted average IV estimator is constructed by first estimating the local average treatment effect (LATE) for each subgroup defined by the same participation probability and then taking a weighted average of these LATE estimates with weights determined by the number of compliers within each subgroup. This estimator coincides with the nonparametric imputation estimator of conditional LATE in \cite{Frolich2007} with participation probability as the key conditioning variable. However, in contrast, the large population random sampling assumption of \cite{Frolich2007} does not apply to online advertising data, so we propose new inference procedures based on design-based uncertainty (discussed below).

Our inference procedure is related to the work of \cite{AbadieAtheyImbensWooldridge2020} and others, which illustrates the difference between sampling-based and design-based inference. Design-based inference focuses on causal parameters defined over the sample and takes into account of the uncertainty from treatment assignment only and treats the units in the data as fixed; while sampling-based inference focuses on causal parameters defined over a hypothetical super-population and takes into account of both uncertainty in treatment assignment and uncertainty in sampling of units. Because the super-population of the potentially non-stationary online traffic is not well-defined and the quantities of primary business interest are causal effects for the realized traffic rather than of some hypothetical super-population, we take a designed-based inference perspective to quantify the uncertainty of the (sample) local average treatment effects we measure. Following \cite{AbadieAtheyImbensWooldridge2020}, this is precisely an example of a scenario where the perspective of design-based uncertainty is valuable. 

Finally, our inference task is related to \cite{hong_inference_2020} which measures the uncertainty of local average treatment effects with a stratified experiment. Our setting can be understood as a complex, \textit{sequential} stratified experiment, where the assignment probability is not a priori fixed, and is driven by the throttling algorithm. We account for the uncertainty of assignment probability through a novel bootstrap procedure we introduce that incorporates the knowledge of the throttling algorithm. We show that this bootstrap procedure generates a conservative estimate of the standard error.

\section{Setup}

Consider an advertiser running a campaign on a platform that serves ads through auctions. 
The advertiser specifies the targeting criteria, campaign duration $T$, bid per impression, and a limited budget. 
Due to this limited budget, the advertiser often cannot reach all eligible impressions that meet the targeting criteria. 
To help the advertiser manage the budget, the advertising platform implements a throttling algorithm $\cA$ that reduces the campaign's probability to participate in auctions for impressions when the budget runs low. 

Refer the $i$th ad auction that meets the campaign's targeting criteria starting from the beginning of the ad campaign as auction $i$. 
For each auction, we observe 
\[
W_i=(p_i, Z_i, D_i, Y_i, X_i, E_i),
\] 
in which $p_i$ is the campaign's participation probability in this auction $i$, $Z_i$ is the binary indicator of whether the advertiser participated in this auction, $D_i$ is referred to as ad exposure or auction outcome, and is the binary indicator of whether the advertiser won the auction and thus displayed its ad to the customer, $Y_i$ is the outcome of interest, e.g. click or conversion, 
\footnote{One challenge is that sometimes advertisers have multiple opportunities to display ads to customers, such that multiple ad impressions are associated with a conversion. For illustration purposes, we first abstract away from the case of multiple ad exposures, and consider the case when advertisers have a single opportunity to display ads to users. In Section \ref{sec:multiple_exposure} we discuss how our method handles situations when advertisers have multiple opportunities to serve ads to the same user.}, 
$X_i$ contains auction specific pretreatment side information like ads quality score and bid, 
and $E_i$ contains additional post-treatment auction information like auction expenditure. 
The throttling algorithm generates a participation probability $p_i$ for auction $i$ based on the available logged information $H_i := (W_j, j=1, \ldots, i-1, X_i)$, which contains auction $i$'s pretreatment side information $X_i$ and all historical data up to auction $i-1$, $(W_j, j=1, \ldots, i-1)$. 
Formally, we have
\[
p_i = \cA(H_i).
\]

To formally discuss the causal inference problem we are interested in, we use the following potential outcome notation. 
Denote the potential ad exposure under auction participation as $D_i(1)$. 
The potential ad exposure under non-participation $D_i(0)=0$ is always equal to zero. 
In other words, we have one-sided compliance. 
The observed ad exposure is $D_i = D_i(Z_i)$. 
Similarly, we denote the post-treatment side information under participation and non-participation as $E_i(1)$ and $E_i(0)$. 
The observed post-treatment side information is $E_i=E_i(Z_i)$. 
Denote the potential outcome of interest under ad exposure and under no ad exposure as $Y_i(1)$ and $Y_i(0)$. 
The observed outcome of interest is $Y_i = Y_i(D_i)$. 
In summary, auction $i$ has the following potential variables,
\[
V_i = (D_i(0), D_i(1), Y_i(0), Y_i(1), X_i, E_i(0), E_i(1)).
\]
All these potential variables are treated as pretreatment characteristics invariant to auction participation.

Let $\cI$ be the set of all auctions that satisfy the campaign's targeting criteria. We show that the platform can measure a local average treatment effect (LATE) of ad exposure defined as
\begin{equation}\label{eqn:objective}
\tau_{\LATE} := \frac{\sum_{i \in \cI} D_i (1) \left[Y_{i}(1)-Y_{i}(0)\right]}{\sum_{i\in \cI} D_i (1)}.
\end{equation}

\subsection*{Interpretation}
$\tau_{\LATE}$ is the average treatment effect of ad exposure on the outcome of interest $Y$ over all compliers, which are all auctions that the advertiser would win if she participates. This estimate is numerically equivalent to the average treatment effect on the treated (ATT) in standard advertising experiments that randomly assigns auction participation with equal probability. 

LATE is different from ATT in our setting, because different auctions have different participation probability due to throttling. This LATE parameter informs the advertiser what the ROI and treatment effect would be, if the campaign has sufficient budget so that it could participate in all auctions. This parameter is structural in the sense it is invariant to the realized participation probability and the actual budget of the campaign. Therefore, if future customers are similar to the current customers, this LATE parameter can better inform the advertiser of future campaign budget allocation. For example, consider a campaign that has spent  90\% of its budget in the morning, and cannot reach all the afternoon customers. If the morning customer has a low ROI of 1\% and the afternoon customer has a high ROI of 50\%, then focusing on ATT, which gives the morning customer a weight of 90\%, may underestimate the potential ROI of the campaign. The advertiser may arrive at the wrong conclusion that they should not allocate budget to run such a campaign in the future. In contrast, focusing on LATE will weigh the morning and afternoon customers not based on how many are actually treated in the past campaign, but based on how many could potentially be treated. Therefore, in the presence of budget limitation, LATE could better inform advertisers of its future budget allocation decisions when future customers are similar to past customers.


%


%



\subsection{Algorithm Assumptions}
%
%
We outline a class of throttling algorithms that enable identification of the LATE. These algorithms satisfy the following assumptions:
\begin{assumption}[Probabilistic Throttling and Conditional Random Assignment of Auction Participation]\label{assumption: unconfoundedness}
The platform uses a throttling algorithm $\cA$ and logged information $H_i$ to \textbf{randomly assign} the focal advertiser into participating auction $i$ with probability $p_i$. Formally,
\[
\cA(H_{i}) = p_i := P(Z_i = 1| H_{i}) = P(Z_i = 1| D_i(0), D_i(1), Y_i(0), Y_i(1), E_i(0), E_i(1), H_{i}).
\]
\end{assumption}
The second equality defines the auction participation probability. 
The first equality says the participation probability is generated from the algorithm $\cA$ and logged information $H_i$. 
The third equality says auction participation $Z_i$ is conditionally randomly assigned given $H_i$. 
In particular, $Z_i$ is conditionally independent from the potential variables.

As a simple example, the vanilla probability throttling algorithm in \cite{karande_optimizing_2013}, which sets the participation probability equal to the ratio of the remaining budget and the remaining maximum spending rate for the rest of the campaign, satisfies assumption \ref{assumption: unconfoundedness}. More intelligent throttling algorithms setting the participation probability adaptively using history and side information also satisfy this assumption.

One natural implication of assumption \ref{assumption: unconfoundedness} is,
\[
P(Z_i = 1| D_i(0), D_i(1), Y_i(0), Y_i(1), E_i(0), E_i(1), p_i) 
= P(Z_i = 1| p_i) = p_i.
\]
Namely, $Z_i$ is conditionally unconfounded given participation probability $p_i$. From a super-poluation perspective, this implication is in the Dawid conditional independence notation:
\begin{implication}[Conditional Random Assignment of Auction Participation]
\label{implication:instrumentvalidity}
\begin{equation}
\left(D_i(0), D_i(1), Y_i(0), Y_i(1), E_i(0), E_i(1)\right) \perp Z_i|p_i.
\label{eq:unconfoundedness}
\end{equation}
\end{implication}

We consider condition \eqref{eq:unconfoundedness} as a novel but pervasive condition under which we can measure the effect of real time bid advertising. The key is to leverage the induced variation in auction participation. In standard observational settings, auction participation is not random marginally and is not a valid (marginal) instrumental variable, because advertiser's targeting criteria often leads to positive correlation between the auction participation and the treatment effect. 
In our setting, participation is not random marginally either, because participation probability $p_i$ is a function of information $H_i$, which may be correlated with the potential outcomes. 
For example, if customer arrivals are serially correlated, then logged information $H_i$ may both affect the current participation probability and be correlated with the current customer's potential outcomes. 
Therefore, participation is not a valid instrumental variable marginally. 
However, \eqref{eq:unconfoundedness} says that after conditioning on the participation probability, participation becomes independent of the potential outcomes due to random assignment. 
\begin{assumption}[Overlap]
\label{assumption: overlap}
There exists some $\eta > 0$, such that for any auction $i \in \cI$, the participation probability $p_i$ generated by the algorithm satisfies $\eta < p_{i} < 1 - \eta$.
\end{assumption}
This assumption implies that the throttling algorithm ensures the participation probabilities are bounded away from zero or one, which allows us to observe sufficient variation in auction participation. 

An alternative approach is the following. 
If we do not assume assumption \ref{assumption: overlap} holds, we can prespecify an $\eta > 0$ and focus on estimating the causal effects over the subpopulation whose observed participation probability lies in $[\eta, 1-\eta]$. 
In this sense, assumption \ref{assumption: overlap} is not essential and is mainly meant to simplify our exposition.
\begin{assumption}[Finite Support]
\label{assumption: finite support}
$p_{i}$s are discrete variables with finite support. 
\end{assumption}
As suggested by assumption \ref{assumption: unconfoundedness}, participation is only unconfounded within the set of auctions with the same participation probability. 
Assumption \ref{assumption: finite support} ensures enough number of auctions share the same participation probability. 
Similar to the propensity score matching literature \citep{abadie_matching_2016}, this finite support assumption can be relaxed to allow for continuous support. 
If we do not assume assumption \ref{assumption: finite support} holds, we can instead partition the participation probability into multiple discrete buckets and then construct estimators using the discretized probabilities. 
In this sense, Assumption \ref{assumption: finite support} is also not essential and is meant to simplify our exposition. Moreover, algorithms in practice often update participation probabilities only at discrete time intervals and therefore satisfy Assumption \ref{assumption: finite support}.



\subsection{Connection to Instrumental Variable}
Auction participation is not a valid instrumental variable marginally. 
However, after conditioning on the participation probability, participation becomes independent of the potential outcomes due to random assignment. 
Therefore, the auction participation can be thought of as a \textbf{conditional} instrumental variable for estimating the advertising effects:
\begin{implication}[Conditional IV]
\label{obs:conditional_iv}
For a subpopulation defined by $\cI_p = \{i \in \cI: p_i = p\}$ with $p\in (0, 1)$, the auction participation $Z_i$ is an instrument variable for estimating the LATE over this subpopulation
\begin{equation}
\label{eq: late p}
\tau_{\LATE, p} = \frac{\sum_{i \in \cI_p} D_i(1) [Y_i(1) - Y_i(0)]}{\sum_{i \in \cI_p} D_i(1)}.
\end{equation}
\end{implication}
Assumption \ref{assumption: unconfoundedness} ensures that $Z_{i}$ satisfies conditional unconfoundedness. 
Because the advertiser cannot win auctions without participation, the monotonicity condition is also satisfied. 
Moreover, the exclusion restriction is automatically satisfied, since the outcome of interest does not depend on auction participation given the auction outcome.

\subsection{Connection to Propensity Scores}
Auction participation probability $p_i$ is similar to but different from a propensity score. 
The key difference is that the propensity score, $P(D_i=1|H_i)$, summarizes the probability of treatment received, but the auction participation probability, $p_i=P(Z_i=1|H_i)$, is the probability of the intent-to-treat. The advertiser may not be explicitly interested in measuring the treatment effect of participating in auctions, but directly interested in the effect of winning auctions to show ads. 
Because participating in auctions is not equivalent to winning auctions, we need to account for the imperfect compliance when measuring the effect of advertising.

\section{Estimation}\label{sec:estimation}
Under Assumptions 1-3, we show that $\tau_{\LATE}$ can be estimated using the following procedure:
\begin{enumerate}
\item Partition the set of auctions $\cI$ based on the participation probability $p$. 
\item Estimate $\tau_{\LATE, p}$ of each partition by treating the participation $Z$ as an instrumental variable. 
\item Estimate the number of compliers in each partition. 
\item Average the $\tau_{\LATE, p}$s of all partitions based on estimated number of compliers in each partition.
\end{enumerate}
Let $\cI_{p, 1} = \{i\in \cI| Z_i=1\}$ and $\cI_{p, 0} = \{i\in \cI| Z_i=0\}$ be the subset of auctions in $\cI_p$ with participation and with nonparticipation, 
$n_p = |\cI_p|$, $n_{p, 1} = |\cI_{p, 1}|$, $n_{p, 0} = |\cI_{p, 0}|$ be the number of auctions in $\cI_p, \cI_{p, 1}, \cI_{p, 0}$, and $\cP$ be the set of all possible $p$s. 
$\tau_{\LATE, p}$ defined in equation \eqref{eq: late p} can be estimated by the standard LATE estimator \citep{imbens_causal_2015}:
\begin{equation}
\hat{\tau}_{\LATE, p} = \frac{\ITThat_{Y, p}}{\ITThat_{D, p}},
\end{equation}
in which 
\begin{align*}
\ITThat_{Y, p} & = \frac{1}{n_{p, 1}}\sum_{i\in \cI_{p, 1}} Y_i - \frac{1}{n_{p, 0}} \sum_{i\in \cI_{p, 0}} Y_i\\
\ITThat_{D, p} & = \frac{1}{n_{p, 1}} \sum_{i\in \cI_{p, 1}} D_i
\end{align*}
are the estimates of intent-to-treat effect on the outcome 
$\ITT_{Y, p} = \frac{1}{n_{p}}\sum_{i\in \cI_{p}} [Y_i(1) -  Y_i(0)]$
and on the treatment 
$\ITT_{D, p} = \frac{1}{n_{p}} \sum_{i\in \cI_{p}} D_i(1)$.

The number of compliers in each partition, $n_{p, co}  = \sum_{i \in \cI_p} D_i(1)$, can be estimated by
\[
\hat{n}_{p, co} = n_p \cdot \frac{\sum_{i \in \cI_{p, 1}} D_i}{n_{p, 1}}.
\]
\begin{prop}
The total LATE $\tau_{\LATE}$ defined in equation \eqref{eqn:objective} can be estimated as a weighted average of $\hat{\tau}_{p}$,
\end{prop}

\begin{equation}
\label{eqn:weighted}
\hat{\tau}_{\LATE} =  \sum_{p \in \cP}\hat{w}_p \cdot \hat{\tau}_{\LATE, p} ,
\end{equation}
where the weight,
\[
\hat{w}_p = \frac{\hat{n}_{p, co}}{\sum_{q \in \cP} \hat{n}_{q, co}}
\]
is the ratio between the estimated number of compliers in partition $\cI_p$ and the estimated number of compliers in the whole population $\cI$. 

\begin{proof}Rewrite equation \eqref{eqn:objective} as follows,
\begin{align*}
\tau_{\LATE} 
& =  \frac{\sum_{i \in \cI} D_i (1) \left[Y_{i}(1)-Y_{i}(0)\right]}{\sum_{i\in \cI} D_i (1)}\\
& = \sum_{p \in \cP} \left\{ \frac{\sum_{i \in \cI_p} D_i(1) \left[Y_{i}(1)-Y_{i}(0)\right]}{\sum_{i\in \cI}D_i(1)}\right\}\\
& = \sum_{p \in \cP} \left\{ \frac{\sum_{i\in \cI_p}D_i(1)}{\sum_{i\in \cI}D_i(1) } \cdot \frac{\sum_{i \in \cI_p} D_i(1)\left[Y_{i}(1)-Y_{i}(0)\right]}{\sum_{i\in \cI_p}D_i(1)}\right\}\\
& = \sum_{p \in \cP} w_p \cdot  \tau_{\LATE, p},
\end{align*}
where the weight $w_p = \frac{\sum_{i\in \cI_p} D_i(1)}{\sum_{i\in \cI}D_i(1)}$ is the number of compliers in group $\cI_p$ divided by the total number of compliers in the whole population $\cI$. 
Replacing $w_p$ and $\tau_{\LATE, p}$ by their estimates $\hat{w}_p$ and $\hat{\tau}_{\LATE, p}$ gives us the $\hat{\tau}_{\LATE}$.
\end{proof}

To build intuition, we can rewrite the above estimator equivalently as,
\begin{equation}
\hat{\tau}_{\LATE} = \frac{\sum_{p \in \cP} n_p \cdot \ITThat_{Y, p}}{\sum_{p \in \cP} n_p \cdot \ITThat_{D, p}}.
\end{equation}
Intuitively, the numerator in the equation above estimates the total intent-to-treat effect if the advertiser were to participate in all the auctions. The denominator estimates the total number of of compliers in the sample, which can be interpreted as the total number of auctions that the advertiser could have won if the advertiser were to participate in all the auctions.

\section{Simulation}\label{sec:simulation}
To demonstrate how our estimator works, we consider a throttling algorithm that updates the participation probability for the focal advertiser every $5$ minutes in a second-price auction. After every $5$-minute interval $t$, the platform calculates the average expenditure per participation $e_t$ of the recent interval, the remaining budget $B_t$, and the number of expected remaining auctions $N_t$. The platform calculates a score $r_t$ using these numbers:
$$
r_{t} = \frac{B_t/e_t}{N_t}
$$
where $\frac{B_t}{e_t}$ is the expected number of auctions that can be participated without violating the campaign budget constraint. Intuitively, when this number is low relative to the number of remaining auctions, the participation probability in the next interval should be low. Assume the participation probability to be a discrete function of such score: 
$$
p_{t+1} = \begin{cases} 
      0.9 & r_{t} \in [0.9, \infty) \\
      0.7 & r_{t} \in [0.7, 0.9)\\
      0.5 & r_{t} \in [0.5, 0.7)\\
      0.3 & r_{t} \in [0.3, 0.5)\\
   \end{cases}
$$

Consider an advertiser that runs a 24-hour campaign with a budget of $\$10,000$ and a bid of $\$5$ per impression. Customers arrival follow a Poisson process at a rate of $10,000$ per day, with two unobserved types, high ($H$) and low ($L$). Amongst $H$ types, $40\%$ will purchase even without ads, and an additional $40\%$ will purchase with ads. H typs are also attractive to competing advertisers, whose highest competing bid is drawn from a uniform distribution of $[4, 7]$. Amongst L types, $10\%$ will purchase without ads, and an additional $10\%$ will purchase after seeing the ads. They are less attractive to competing advertisers, whose highest competing bid is drawn from a uniform distribution of $[1, 6]$. Customer types have a serial correlation of $0.99$ , which is meant to model typical online traffic where similar customers tend to arrive around similar time.

\begin{figure}[H]
\centering
\includegraphics[scale=0.7]{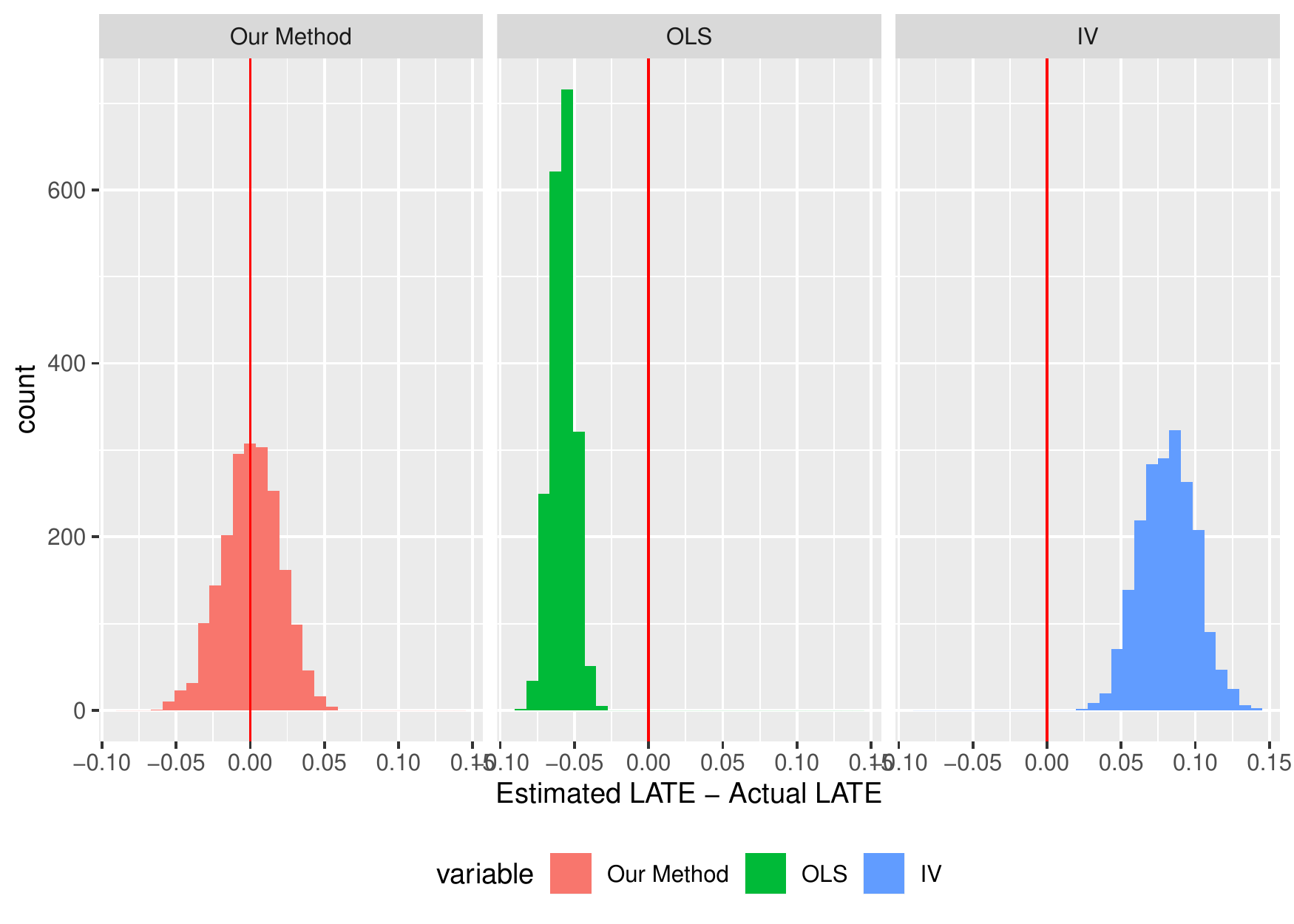}
\caption{Distribution of difference between $\tau_{\LATE}$ and estimates: $\hat{\tau}_{\LATE}$ vs OLS vs IV}\label{fig:iv_comparison}
\end{figure}

We simulate such campaign for $1,000$ times, and calculate the difference between $\tau_{\LATE}$ with estimates using our estimator $\hat{\tau}_{\LATE}$ in Equation \ref{eqn:weighted}, as well as standard OLS and IV estimators: 

Table \ref{tab:performance_difference} demonstrates the bias of these estimators. Figure \ref{fig:iv_comparison} shows the distribution of such differences.

\begin{table}[htbp!]
\caption{Method Comparison}\label{tab:performance_difference}
\centering

\begin{tabular}{l|c|rrr}
\toprule
Method & Estimator & Mean & Bias & RMSE\\
\midrule
$\tau_{\LATE}$ & & 0.2245 & 0.0000 & 0.0000\\
$\hat{\tau}_{\LATE}$ & Equation \eqref{eqn:weighted}& 0.2251 & 0.0006 & 0.0255\\
OLS & $\hat{E}[Y_i|D_i =1]- \hat{E}[Y_i|D_i = 0]$& 0.1530 & -0.0714 & 0.0718\\
IV & $\frac{\hat{E}[Y_i|Z_i =1]- \hat{E}[Y_i|Z_i = 0]}{\hat{E}[D_i|Z_i =1]- \hat{E}[D_i|Z_i = 0]}$ & 0.2924 & 0.0680 & 0.0703\\
\bottomrule
\end{tabular}

\end{table}

The OLS estimator is biased because it incorrectly assumes winning the auction to be random. Winning the auction is endogenous because the winning probability depends on the customer type, which affects the bid of the competitor. The IV estimator is also biased, because it incorrectly assumes the participation to be random. The participation is endogenous because of serial correlation: past customer types are not only correlated with current customer types but also affect current participation. Appendix \ref{appendix:simulation_detail} provides a more detailed description of the simulation process.

\section{Inference}

Quantifying the uncertainty of our estimate is challenging because customer arrivals and advertiser auction participations can all be serially correlated and even nonstationary. 
Since it is hard to retrospectively characterize the super-population from which impressions are drawn and in-sample treatment effect is of primary business interest, we will treat the sample in the data as fixed (equivalently conditional on the data) and take a \emph{design-based inference} perspective to quantify uncertainty of our estimated LATE following \cite{AbadieAtheyImbensWooldridge2020}: given the set of units in the sample, how would the estimated LATE be different from the actual LATE, if the auction participations were assigned differently in a thought experiment.
Formally, this design-based uncertainty is defined as:
$$
\var(\widehat{\tau}_{\LATE}|\cI)
$$
where $\cI$ is the observed sample for which we are interested in evaluating the advertising effectiveness. 

 Because auction participations are assigned by the throttling algorithm, the distribution of the estimator and its inference also depends on the algorithm. 
 We focus on a class of throttling algorithms that update the participation probability for the focal advertiser every $M$ minutes. We demonstrate how to conduct design-based inference through bootstrap when researchers have access to the exact throttling algorithm, which is the situation for platforms that have perfect information over how the participation probability is generated.


To formally define the throttling algorithm, index the time interval by $t \in \{0, 1, \ldots, T\}$, where each interval last $M$ minutes. Let $\cI_t$ be the set of auctions that belong to interval $t$, and $p_t$ be the participation probability for all auctions in $i \in \cI_t$. The throttling algorithm $\cA$ maps past history $H_t$ to the participation probability for the next interval:
$$
p_{t+1} = \cA(H_t),
$$
where $H_t = \{W_i, i \in \cI_0 \cup \cI_1 ... \cup \cI_t\}$ is history observed by the platform after the end of interval $t$. 
Figure \ref{fig:DGP} gives an example of the DAG of the data generating process. 

\begin{figure}[htbp!]
\caption{Data generating process for the throttling algorithm}\label{fig:DGP}
\centering
\resizebox{.7\textwidth}{!}{%
\begin{tikzpicture}[->,>=stealth',shorten >=1pt,auto,node distance=3cm,
                    semithick, label distance=20pt]
  \tikzstyle{every state}=[fill=white, draw=black, text=black, shape = rectangle] 

  \node[initial, state, label=above:{$t=0$}] (p0)                    {\text{Initial} $p_0$};
  \node[state]                               (Z0) [below of=p0]      {$\{W_i, i \in \cI_0\}$};
  \node[state]                               (h0) [below of=Z0]      {$H_0$} ; 
  \node[state, label=above:{$1$}]            (p1) [right of=p0]      {$p_1 = \cA(H_0)$};
  \node[state]                               (Z1) [below of=p1]      {$\{W_i, i \in \cI_1\}$};
  \node[state]                               (h1) [below of=Z1]      {$H_1$} ; 
  \node[state, label=above:{$2$}]            (p2) [right of=p1]      {$p_2 = \cA(H_1)$};
  \node[state]                               (Z2) [below of=p2]      {$\{W_i, i \in \cI_2\}$};
  \node[state]                               (h2) [below of=Z2]      {$H_2$} ; 
  \node[state, label=above:{$3$}]            (p3) [right of=p2]      {$p_3 = \cA(H_2)$};
  \node[state]                               (Z3) [below of=p3]      {$\{W_i, i \in \cI_3\}$};
  \node[state]                               (h3) [below of=Z3]      {$H_3$} ; 
  \path (p0) edge              node {} (Z0)
        (Z0) edge              node {} (h0)
         (h0) edge              node {} (p1)
         (h0) edge              node {} (h1)
        (p1) edge              node {} (Z1)
        (Z1) edge              node {} (h1)
        (h1) edge              node {} (p2)   
        (h1) edge              node {} (h2)   
        (p2) edge              node {} (Z2)
        (Z2) edge              node {} (h2)            
        (h2) edge              node {} (p3)    
        (h2) edge  		       node {} (h3)   
        (p3) edge              node {} (Z3)
        (Z3) edge              node {} (h3)
        ;
\end{tikzpicture}}
\end{figure}

\subsection{Bootstrap Inference}
Our inference procedure first generates bootstrap samples  that could have been generated under the throttling algorithm. 
A bootstrap sample $\cS^b$ of the campaign can be simulated using the bootstrap sampler below. 
Each bootstrap sample $\cS^b$ contains an entire sequence of samples from $t=0$ to $t=T$. 
We use $\cS_{t, 1} = \{W_i, i\in I_{t, 1}\}$ and $\cS_{t, 0} = \{W_i, i\in I_{t, 0}\}$ to denote data in period $t$ with participation and with nonparticipation. 
We use $\operatorname{Resample}(S, m)$ to denote $m$ uniform random sample from dataset $S$ with replacement.

\begin{algorithm}[H]
\SetKwComment{Comment}{/* }{ */}
\SetKwInOut{Input}{input}
\SetKwInOut{Output}{output}
\caption{Bootstrap sampler}
\label{algo: bootstrap sampler}
	\Input{Data $\cS=\{W_i, i \in \cI_0 \cup \cdots \cup \cI_T\}$, with $W_i = (p_i, Z_i, D_i, Y_i, X_i, E_i)$}
	\Output{Bootstrap sample $\cS^b$}
	\BlankLine
	$p_0^b \gets p_0$\;
	\For{$t$ in $[0, 1, \ldots, T]$}{
		$n_{t, 1}^b \gets \operatorname{Binomial}(n_t, p_t)$ \Comment*[r]{Draw the total number of participations based on $p_t$}
		$\cS_{t, 1}^b \gets \operatorname{Resample}(\cS_{t, 1}, n_{t, 1}^b)$ \Comment*[r]{Draw $n_{t, 1}^b$ bootstrap samples with participation}
		$\cS_{t, 0}^b \gets \operatorname{Resample}(\cS_{t, 0}, n_{t} - n_{t, 1}^b)$ \Comment*[r]{Draw $n_{t} - n_{t, 1}^b$ bootstrap samples with nonparticipation}
		$\cS_t^b \gets [\cS_{t, 1}, \cS_{t, 0}]$ \Comment*[r]{Combine $\cS_{t, 1}^b$ and $\cS_{t, 0}^b$ as all bootstrap samples in period $t$}
		$H_t^b \gets [H_{t-1}^b, \cS_t^b]$ \Comment*[r]{Combine $H_{t-1}^b$ and $\cS_{t}^b$ as history at the end of period $t$}
		$p_{t+1}^b \gets \cA(H_t^b)$ \Comment*[r]{Calculate participation probability in the next period}
	}
	$\cS^b \gets H_T^b$
\end{algorithm}
One key difference from the standard bootstrap procedure is that participation cannot be directly simulated from the empirical distribution by sampling with replacement. 
Instead, it has to be simulated based on the participation probability that is an output of the throttling algorithm. 
Knowledge of algorithm is key for conducting inference, because how the participation probability is generated/sampled cannot be directly derived from the empirical data.

Another key difference from the standard bootstrap procedure is that in this causal inference set-up, some of the potential outcomes are missing, i.e. we only observe the potential outcome under auction participation or under auction nonparticipation but not both for each impression. 
We address this missing data problem by drawing bootstrap samples with participation and with nonparticipation using separate subpopulations with participation and with nonparticipation respectively. 

After simulating many such bootstrap samples $\cS^b, b=1, \ldots, B$ and generate corresponding bootstrap version of the estimate $\hat{\tau}_{\LATE}^b = \hat{\tau}_{\LATE}(\cS^b)$, we can construct an estimate of the variance $\var(\hat{\tau}_{\LATE})$ as:
\[
\widehat{\var}_{bootstrap}(\hat{\tau}_{\LATE}) 
:= \var(\hat{\tau}_{\LATE}(\cS^b))
= \frac{1}{B-1} \sum_{b=1}^B \left(\hat{\tau}_{\LATE}(\cS^b) - \frac{\sum_{c=1}^B \hat{\tau}_{\LATE}(\cS^c)}{B}\right)^2
\]
This variance estimate is expected to be conservative because 1) bootstrap is generally known to be valid for asymptotic inference under the superpopulation model \cite{hong_inference_2020}, and 2) the standard error estimated under a superpopulation model can be used as a conservative standard error for our sample LATE under a finite population model (\cite{imbens_nonparametric_2003} and \cite{AbadieAtheyImbensWooldridge2020}). 

Given the bootstrap resamples, we can also construct a bootstrap confidence interval using a percentile method:
\[
\operatorname{CI}(\alpha) = [\hat{\tau}_{\LATE} - \hat{\xi}_{1-\alpha/2}, \hat{\tau}_{\LATE} - \hat{\xi}_{\alpha/2}],
\]
in which $\hat{\xi}_{\alpha/2}$ and $\hat{\xi}_{1-\alpha/2}$ are $\alpha/2$-percentile and $1-\alpha/2$-percentile of the empirical distribution $\hat{\tau}_{\LATE}(\cS^b) - \hat{\tau}_{\LATE}, b=1, \ldots, B$.  

\subsection{Inference Based on Asymptotic Normal Approximation}
The bootstrap inference method proposed above works well in general but requires access to the throttling algorithm. 
Here we propose an alternative inference method based on large-sample asymptotic normal approximation. 
This method does not require access to the throttling algorithm and is provably valid under the additional assumption that participation probabilities are a priori fixed. 
Though there is no formal guarantee that the confidence interval based on this method has the correct coverage when the participation probabilities are adaptive, our simulation suggests the method often works well empirically for large campaigns. 
As a result, we believe this is a reasonable inference method in practice. 

Our large-sample approximation is based on the following representation:
\begin{equation*}
\hat{\tau}_{\LATE} = \frac{\sum_{p \in \cP} n_p \cdot \ITThat_{Y, p}}{\sum_{p \in \cP} n_p \cdot \ITThat_{D, p}}.
\end{equation*}
Suppose participation probabilities are fixed. 
Since all the $\ITThat_{Y, p}$s and $\ITThat_{D, p}$s are sample means or differences of sample means, each of them are asymptotically normal. 
As a smooth function of sample means, $\hat{\tau}_{\LATE}$ is also going to be asymptotically normal. 
More precisely,
\begin{align*}
\sqrt{n}\left(\hat{\tau}_{\LATE} - \tau_{\LATE}\right)
\overset{d}{\rightarrow} 
\Normal\left(0, \Omega\right),
\end{align*}
where $\Omega$ is the asymptotic variance. 
Proposition \ref{prop: asymptotic normality} in appendix \ref{sec:fixed_prob} makes this statement rigorous and presents a proof outline based on delta-method and central limit theorem. 
Since variance of the estimator can be approximated by
$
\var\left(\hat{\tau}_{\LATE}\right)
\approx \sum_{p\in \cP} w_p^2 \cdot  \var\left(\frac{\ITThat_{Y, p} - \ITThat_{D, p} \cdot \tau_{\LATE}}{\ITT_{D, p}}\right),
$
intuitively one can can estimate the variance by a plug-in estimator
\begin{align*}
\widehat{\var}\left(\hat{\tau}_{\LATE}\right)
& = \sum_{p\in \cP} \hat{w}_p^2 \cdot  \widehat{\var}\left(\frac{\ITThat_{Y, p} - \ITThat_{D, p} \cdot \tau_{\LATE}}{\ITT_{D, p}}\right). 
\end{align*}
Equation \eqref{eq: variance estimator} in Appendix \ref{sec:fixed_prob} presents a full specification of the variance estimator. 

This variance estimator may slightly overstate the variance, because the standard sampling-based variance estimators of intent-to-treat effects tend to be conservative for finite-sample design-based inference \cite{AbadieAtheyImbensWooldridge2020}.



Whether the confidence interval based on this estimated variance and asymptotic normality works or not depends on how participation probabilities vary in practice and whether we have large samples or not for each $p$. 
We find that this confidence interval often works particularly well for large advertising campaigns. 
The large sample size and relatively smooth change in participation probabilities of large campaigns fits the assumptions of this method well. 

Table \ref{tab:coverage_difference} shows the coverage using this asymptotic-normality-based approach and the bootstrap approaches. Given our throttling algorithm in Section \ref{sec:simulation}, both methods provide a conservative estimate of the true variance.

\begin{table}[H]
\caption{Simulation Results With Coverage For Nominal $95\%$ Confidence Intervals}\label{tab:coverage_difference}
\centering
\begin{tabular}{lccc}
\toprule
& $\var(\widehat{\tau}_{\LATE})^{1/2}$  & $\widehat{\var}(\hat{\tau}_{\LATE})^{1/2}$ & $\widehat{\var}_{bootstrap}(\hat{\tau}_{\LATE})^{1/2}$\\
\midrule
Standard Error  & 0.0195 & 0.0205 & 0.0245 \\
$95\%$ Coverage &  & $96.1\%$ & $98.6\%$ \\
\bottomrule
\end{tabular}
\end{table}

The magnitude of under/over coverage of the asymptotic-normality-based inference method depends on the exact throttling algorithm and the serial correlation of customer types. 
Appendix \ref{sec:fixed_prob} provides some intuition on why there may be over coverage in our simulation exercise.


%

%
%
%
%
%
%
%

\section{Empirical Application}

\begin{figure}[H]
\centering
\includegraphics[scale=0.7]{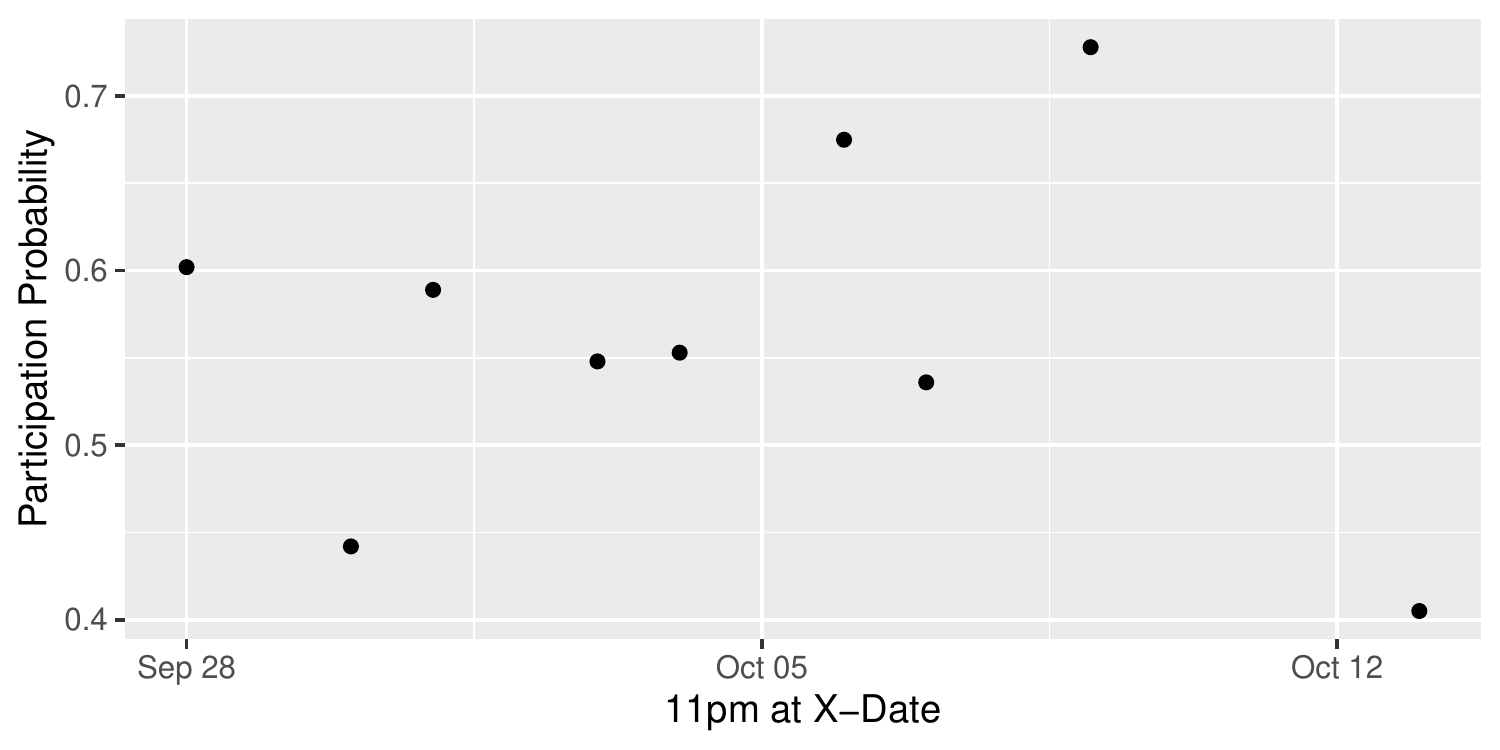}
\caption{Participation Probability Vs Time}\label{fig:jd_participation_prob}
\end{figure}

We apply our method to a campaign run on a large e-commerce advertising platform. The platform sells ads through auction and uses probabilistic throttling as a budget control mechanism. The campaign usually has low budget near the end of the day, triggering the throttling mechanism. We analyze the campaign data from $11:00$pm-$11:59$pm of each day, for a total of 9 days. Figure \ref{fig:jd_participation_prob} shows how the participation probability is different at $11$pm across different days. 

Our outcome of interest $\{Y_i(1), Y_i(0)\}$ is whether the customer has visited the product page related to the advertiser, regardless of whether it is through ad or organic search. For each hour, we observe all auctions participated by the focal campaign, as well as a sample of auctions not participated by the focal campaign. This nonparticipation is primarily driven by either throttling or not meeting the targeting criteria. Ideally our method can be best applied if the platform has logged whether the auction meets the targeting criteria, such that we can observe all auctions that the focal campaign could have participated in the absence of throttling. Unfortunately, this information is usually not logged, because the value of this information and how it can be applied has not been clearly documented prior to our paper. To demonstrate the value of our approach, we impute these throttling-induced non-participation based on whether the focal campaign's top-20 competitors have participated and how much these top competitors bid. \footnote{The top competitors are measured by how frequently they appear in the auctions participated by the focal advertiser.} If a set of auctions have exactly the same top competitors and bids, but only some auctions are participated by the focal campaign, then we label the remaining auctions as throttling-induced non-participation. \footnote{This prediction procedure is most similar to matching, but there are several differences: 1) the source of variation is directly known by researchers because of throttling; 2) instead of predicting units that could have received the treatment, the ad exposure, we are predicting the units that could have received the instrument, the auction participation; 3) the expected number of nonparticipated units $n_{t,0}$ to be matched to the participated units is directly informed by the throttling algorithm: $n_{t,0} = \frac{n_{t,1} \cdot (1 - p_t)}{p_{t}}$. In our empirical setting, we cannot find all of these nonparticipated units, because we only observe a sub-sample of auctions not participated by the focal campaign, but not all of them. To ensure our estimate is valid, we weight the matched non-participated units so their weight add up to $n_{t, 0}$.} We then apply our estimation method on this subset of auctions where we can match participated auctions with non-participated auctions. In practice, to avoid this additional imputation step that may introduce additional uncertainty, we recommend the platform to directly log auctions that the advertisers were throttled out. 

 Table \ref{tab:empirical_estimates} compares our method with standard observational methods that do not account for the reasons behind nonparticipation. Column (1) shows the OLS estimate that incorrectly assumes the ad exposure to be random. Column (2) shows the IV estimate that incorrectly assumes the auction participation to be a valid IV. Column (3) shows our method in Section \ref{sec:estimation} that leverages the knowledge of the throttling algorithm, which allows us to use participation as a conditional IV and excludes the auctions that the advertiser would never have participated in without the budget constraint.


\begin{table}[H] \centering 
  \caption{Estimates of LATE} 
  \label{tab:empirical_estimates} 
\begin{tabular}{@{\extracolsep{5pt}}lccc} 
\\[-1.8ex]\hline 
\hline \\[-1.8ex] 
 & \multicolumn{3}{c}{\textit{Dependent variable: }} \\ 
\cline{2-4} 
\\[-1.8ex] & \multicolumn{3}{c}{Product Page View} \\ 
\\[-1.8ex] & OLS & IV & Our Method ($\hat{\tau}_{\LATE}$) \\ 
\hline \\[-1.8ex] 
 Treatment Effect &  0.019$^{***}$ & 0.044$^{***}$ &  0.011$^{***}$  \\ 
  & (0.0004) & (0.0021) & (0.0027) \\ 
\hline \\
Observations & 1,257,793 & 1,257,793  & 995,311 \\ 
\hline 
\hline \\[-1.8ex] 
\textit{Note:}  & \multicolumn{3}{r}{$^{*}$p$<$0.1; $^{**}$p$<$0.05; $^{***}$p$<$0.01} \\ 
\end{tabular} 
\end{table} 

Compared to our estimate, the OLS and the incorrect IV both overestimate the treatment effect, respectively by 72\% and 300\%. 
This result is consistent with related literature that compares standard observational methods with experimental data (\cite{blake_consumer_2015} and \cite{gordon_comparison_2019}, and \cite{gordon_close_2022}), which finds that standard observational methods tend to over-state the benefit of advertising. To benchmark the degree of overestimation, we follow \cite{gordon_close_2022} to report the conversion lift of the estimated treatment effect, defined as 

\begin{align*}
  l & = \frac{\text{Conversion rate due to ad exposure}}{\text{Baseline conversion rate if customers are not exposed to ads}} \\
  & = \frac{\tau}{E[Y_i|D_i = 1] - \tau}
\end{align*}

Both $E[Y_i|D_i = 1]$ and $\tau$ can be estimated by the data, we report the estimated baseline conversion rate and the conversion lift in Table \ref{tab:conversion_lift}. 

\begin{table}[htbp!]
\caption{Method Comparison}\label{tab:conversion_lift}
\centering

\begin{tabular}{lrrr}
\toprule
 & OLS & IV & Our Method\\
\midrule
Treatment effect & 0.019 & 0.044 & 0.011 \\
Baseline conversion & 0.003 & -0.022 & 0.010\\
Conversion lift & 666\% & NA & 110\%\\
\bottomrule
\end{tabular}
\end{table}

The conversion lift for the IV estimate cannot be calculated because the estimated baseline conversion is negative, which is impossible because the minimum conversion rate is $0$. This negative baseline conversion directly rejects that participation is a valid instrument. The conversion lift estimated using OLS, 600\%, is much higher than the conversion lift estimated using our method, 110\%, suggesting OLS may underestimate the baseline conversion rate. This underestimation is due to the fact that ad exposure is not random and may be positively correlated with the potential outcome, such that $E[Y_i(0)|D_i = 0] < E[Y_i(0)|D_i = 1]$. 


%

\section{Conclusion}

We demonstrate that probabilistic auction throttling, a budget control mechanism widely used by advertising platforms, induces pervasive quasi-experimental variation in auction participation. To leverage this variation for measuring the effectiveness of digital advertising, we propose a weighted IV estimator, where auction participation is a conditional instrumental variable. To account for the uncertainty in this novel setting, we also develop a bootstrap procedure that leverages the knowledge of the throttling algorithm to account for design-based uncertainty. 

We use simulation to show that our method causally identifies the LATE. We then apply our method on an advertising campaign from an e-commerce advertising platform that used a probabilistic throttling algorithm. We show that our LATE estimate is significantly different from estimates obtained using other standard observational methods including OLS and IV. Our estimated conversion lift is 110\%, a more plausible number than 600\%, the conversion lifts estimated using naive observational methods. These results suggest that standard observational methods generates biases in our context, and highlights the need to leverage exogenous variation generated by the throttling algorithm.  


\section{Future Extensions: Multiple Ad Exposure}\label{sec:multiple_exposure}

Because our primary goal is to illustrate the value of the variation induced by probabilistic throttling, we mainly demonstrated the case where the campaign has one advertising opportunity for each user. By an ``advertising opportunity,'' we mean an auction that the advertiser could \textit{potentially} participate in, henceforth referred to as a ``potential auction.'' The procedure we have outlined can be extended easily to allow for cases when the campaign has multiple opportunities to advertise to each user. 

The idea is to first divide users into distinct groups based on the number of \textit{potential} auctions that are associated with each user who meets the focal advertiser's targeting criteria. For example, group 1 would comprise users with exactly 1 potential auction, and group 2 would comprise users with exactly 2 potential auctions, and so forth. The key idea is to stratify users based on the potential, rather than the actual, number of auctions that the advertiser could have participated in for that user. This stratification could be thought of as a specific form of principle stratification (\cite{frangakis_principal_2002}), except that the potential number of auctions may be directly observed in the data. For example, if the platform has logged all eligible auctions that meet the advertiser's targeting criteria, then we directly observe the set of all potential auctions. The number of potential auctions for a given user can be directly calculated by summing up the number of potential auctions associated with the user. After this stratification, we can apply our estimator separately to each group to estimate a group-specific LATE. The advertiser could then utilize these group-specific LATE-s as is. Or they could combine them by taking a weighted average in order to generate an overall LATE across all groups.  Intuitively, this procedure would allow advertisers to compare users associated with the same number of advertising opportunities and participation probabilities, but are exposed to different numbers and types of ads due to throttling and budget limitation. It has important managerial implication for advertisers, because they can better understand the ROI of different types of ads, how they interact with each other, and the optimal intensity of ads.

\newpage

\bibliographystyle{apalike}
\bibliography{Reference.bib,AuctionThrottling.bib,ExperimentValue.bib} %

\begin{thebibliography}{}

\bibitem[Abadie, 2003]{Abadie2003}
Abadie, A. (2003).
\newblock Semiparametric instrumental variable estimation of treatment response
  models.
\newblock {\em Journal of econometrics}, 113(2):231--263.

\bibitem[Abadie et~al., 2020]{AbadieAtheyImbensWooldridge2020}
Abadie, A., Athey, S., Imbens, G.~W., and Wooldridge, J.~M. (2020).
\newblock Sampling-based versus design-based uncertainty in regression
  analysis.
\newblock {\em Econometrica}, 88(1):265--296.

\bibitem[Abadie and Imbens, 2016]{abadie_matching_2016}
Abadie, A. and Imbens, G.~W. (2016).
\newblock Matching on the {Estimated} {Propensity} {Score}.
\newblock {\em Econometrica}, 84(2):781--807.

\bibitem[Agarwal et~al., 2014]{agarwal_budget_2014}
Agarwal, D., Ghosh, S., Wei, K., and You, S. (2014).
\newblock Budget pacing for targeted online advertisements at {LinkedIn}.
\newblock In {\em Proceedings of the 20th {ACM} {SIGKDD} international
  conference on {Knowledge} discovery and data mining}, {KDD} '14, pages
  1613--1619, New York, New York, USA. Association for Computing Machinery.

\bibitem[Amin et~al., 2012]{amin2012budget}
Amin, K., Kearns, M., Key, P., and Schwaighofer, A. (2012).
\newblock Budget optimization for sponsored search: censored learning in mdps.
\newblock In {\em Proceedings of the Twenty-Eighth Conference on Uncertainty in
  Artificial Intelligence}, pages 54--63.

\bibitem[Angrist et~al., 1996]{AngristImbensRubin1996}
Angrist, J.~D., Imbens, G.~W., and Rubin, D.~B. (1996).
\newblock Identification of causal effects using instrumental variables.
\newblock {\em Journal of the American statistical Association},
  91(434):444--455.

\bibitem[Balseiro et~al., 2021]{balseiro_budget_2017}
Balseiro, S., Kim, A., Mahdian, M., and Mirrokni, V. (2021).
\newblock Budget-management strategies in repeated auctions.
\newblock {\em Operations Research}, 69(3):859--876.

\bibitem[Blake et~al., 2015]{blake_consumer_2015}
Blake, T., Nosko, C., and Tadelis, S. (2015).
\newblock Consumer {Heterogeneity} and {Paid} {Search} {Effectiveness}: {A}
  {Large}-{Scale} {Field} {Experiment}.
\newblock {\em Econometrica}, 83(1):155--174.
\newblock \_eprint: https://onlinelibrary.wiley.com/doi/pdf/10.3982/ECTA12423.

\bibitem[Borgs et~al., 2007]{borgs2007dynamics}
Borgs, C., Chayes, J., Immorlica, N., Jain, K., Etesami, O., and Mahdian, M.
  (2007).
\newblock Dynamics of bid optimization in online advertisement auctions.
\newblock In {\em Proceedings of the 16th international conference on World
  Wide Web}, pages 531--540.

\bibitem[Charles et~al., 2013]{charles2013budget}
Charles, D., Chakrabarty, D., Chickering, M., Devanur, N.~R., and Wang, L.
  (2013).
\newblock Budget smoothing for internet ad auctions: a game theoretic approach.
\newblock In {\em Proceedings of the fourteenth ACM conference on Electronic
  commerce}, pages 163--180.

\bibitem[Feldman et~al., 2007]{feldman2007budget}
Feldman, J., Muthukrishnan, S., Pal, M., and Stein, C. (2007).
\newblock Budget optimization in search-based advertising auctions.
\newblock In {\em Proceedings of the 8th ACM conference on Electronic
  commerce}, pages 40--49.

\bibitem[Frangakis and Rubin, 2002]{frangakis_principal_2002}
Frangakis, C.~E. and Rubin, D.~B. (2002).
\newblock Principal {Stratification} in {Causal} {Inference}.
\newblock {\em Biometrics}, 58(1):21--29.

\bibitem[Fr{\"o}lich, 2007]{Frolich2007}
Fr{\"o}lich, M. (2007).
\newblock Nonparametric iv estimation of local average treatment effects with
  covariates.
\newblock {\em Journal of Econometrics}, 139(1):35--75.

\bibitem[Geng et~al., 2021]{geng_automated_2021}
Geng, T., Sun, F., Wu, D., Zhou, W., Nair, H., and Lin, Z. (2021).
\newblock Automated {Bidding} and {Budget} {Optimization} for {Performance}
  {Advertising} {Campaigns}.
\newblock {\em SSRN Electronic Journal}.

\bibitem[Gordon et~al., 2021]{gordon2021inefficiencies}
Gordon, B.~R., Jerath, K., Katona, Z., Narayanan, S., Shin, J., and Wilbur,
  K.~C. (2021).
\newblock Inefficiencies in digital advertising markets.
\newblock {\em Journal of Marketing}, 85(1):7--25.

\bibitem[Gordon et~al., 2022]{gordon_close_2022}
Gordon, B.~R., Moakler, R., and Zettelmeyer, F. (2022).
\newblock Close {Enough}? {A} {Large}-{Scale} {Exploration} of
  {Non}-{Experimental} {Approaches} to {Advertising} {Measurement}.
\newblock {\em arXiv:2201.07055 [econ]}.
\newblock arXiv: 2201.07055.

\bibitem[Gordon et~al., 2019]{gordon_comparison_2019}
Gordon, B.~R., Zettelmeyer, F., Bhargava, N., and Chapsky, D. (2019).
\newblock A {Comparison} of {Approaches} to {Advertising} {Measurement}:
  {Evidence} from {Big} {Field} {Experiments} at {Facebook}.
\newblock {\em Marketing Science}, 38(2):193--225.
\newblock Publisher: INFORMS.

\bibitem[Hartmann and Klapper, 2018]{hartmann2018super}
Hartmann, W.~R. and Klapper, D. (2018).
\newblock Super bowl ads.
\newblock {\em Marketing Science}, 37(1):78--96.

\bibitem[Hong et~al., 2020]{hong_inference_2020}
Hong, H., Leung, M.~P., and Li, J. (2020).
\newblock Inference on finite-population treatment effects under limited
  overlap.
\newblock {\em The Econometrics Journal}, 23(1):32--47.

\bibitem[Hong and Nekipelov, 2010]{Hong2010}
Hong, H. and Nekipelov, D. (2010).
\newblock Semiparametric efficiency in nonlinear late models.
\newblock {\em Quantitative Economics}, 1(2):279--304.

\bibitem[Imbens, 2003]{imbens_nonparametric_2003}
Imbens, G.~W. (2003).
\newblock Nonparametric {Estimation} of {Average} {Treatment} {Effects} under
  {Exogeneity}: {A} {Review}.
\newblock Working {Paper} 294, National Bureau of Economic Research.
\newblock Series: Technical Working Paper Series.

\bibitem[Imbens and Angrist, 1994]{ImbensAngrist1994}
Imbens, G.~W. and Angrist, J.~D. (1994).
\newblock Identification and estimation of local average treatment effects.
\newblock {\em Econometrica: Journal of the Econometric Society}, pages
  467--475.

\bibitem[Imbens and Rubin, 2015]{imbens_causal_2015}
Imbens, G.~W. and Rubin, D.~B. (2015).
\newblock {\em Causal {Inference} for {Statistics}, {Social}, and {Biomedical}
  {Sciences}: {An} {Introduction}}.
\newblock Cambridge University Press, 1 edition.

\bibitem[Johnson et~al., 2017]{johnson_ghost_2017}
Johnson, G.~A., Lewis, R.~A., and Nubbemeyer, E.~I. (2017).
\newblock Ghost {Ads}: {Improving} the {Economics} of {Measuring} {Online} {Ad}
  {Effectiveness}.
\newblock {\em Journal of Marketing Research}, 54(6):867--884.

\bibitem[Karande et~al., 2013]{karande_optimizing_2013}
Karande, C., Mehta, A., and Srikant, R. (2013).
\newblock Optimizing budget constrained spend in search advertising.
\newblock In {\em Proceedings of the sixth {ACM} international conference on
  {Web} search and data mining - {WSDM} '13}, page 697, Rome, Italy. ACM Press.

\bibitem[Lewis and Rao, 2015]{lewis_unfavorable_2015}
Lewis, R.~A. and Rao, J.~M. (2015).
\newblock The {Unfavorable} {Economics} of {Measuring} the {Returns} to
  {Advertising}*.
\newblock {\em The Quarterly Journal of Economics}, 130(4):1941--1973.

\bibitem[Muthukrishnan et~al., 2007]{muthukrishnan2007stochastic}
Muthukrishnan, S., Pal, M., and Svitkina, Z. (2007).
\newblock Stochastic models for budget optimization in search-based
  advertising.
\newblock In {\em International Workshop on Web and Internet Economics}, pages
  131--142. Springer.

\bibitem[Muthukrishnan et~al., 2010]{muthukrishnan2010stochastic}
Muthukrishnan, S., P{\'a}l, M., and Svitkina, Z. (2010).
\newblock Stochastic models for budget optimization in search-based
  advertising.
\newblock {\em Algorithmica}, 58(4):1022--1044.

\bibitem[Nuara et~al., 2018]{nuara2018combinatorial}
Nuara, A., Trovo, F., Gatti, N., and Restelli, M. (2018).
\newblock A combinatorial-bandit algorithm for the online joint bid/budget
  optimization of pay-per-click advertising campaigns.
\newblock In {\em Proceedings of the AAAI Conference on Artificial
  Intelligence}.

\bibitem[Sahni, 2015]{Sahni2015}
Sahni, N.~S. (2015).
\newblock Effect of temporal spacing between advertising exposures: Evidence
  from online field experiments.
\newblock {\em Quantitative Marketing and Economics}, 13(3):203--247.

\bibitem[Sahni et~al., 2019]{sahni2019retargeted}
Sahni, N.~S., Narayanan, S., and Kalyanam, K. (2019).
\newblock An experimental investigation of the effects of retargeted
  advertising: The role of frequency and timing.
\newblock {\em Journal of Marketing Research}, 56(3):401--418.

\bibitem[Shapiro, 2018]{shapiro_positive_2018}
Shapiro, B.~T. (2018).
\newblock Positive {Spillovers} and {Free} {Riding} in {Advertising} of
  {Prescription} {Pharmaceuticals}: {The} {Case} of {Antidepressants}.
\newblock {\em Journal of Political Economy}, 126(1):381--437.
\newblock Publisher: The University of Chicago Press.

\bibitem[Shapiro et~al., 2021]{shapiro2021tv}
Shapiro, B.~T., Hitsch, G.~J., and Tuchman, A.~E. (2021).
\newblock Tv advertising effectiveness and profitability: Generalizable results
  from 288 brands.
\newblock {\em Econometrica}, 89(4):1855--1879.

\bibitem[Stephens-Davidowitz et~al., 2017]{stephens2017super}
Stephens-Davidowitz, S., Varian, H., and Smith, M.~D. (2017).
\newblock Super returns to super bowl ads?
\newblock {\em Quantitative Marketing and Economics}, 15(1):1--28.

\bibitem[Zhang et~al., 2012]{zhang_joint_2012}
Zhang, W., Zhang, Y., Gao, B., Yu, Y., Yuan, X., and Liu, T.-Y. (2012).
\newblock Joint optimization of bid and budget allocation in sponsored search.
\newblock In {\em Proceedings of the 18th {ACM} {SIGKDD} international
  conference on {Knowledge} discovery and data mining}, pages 1177--1185.
\newblock 00041.

\bibitem[Zhou et~al., 2008]{chakrabarty2007budget}
Zhou, Y., Chakrabarty, D., and Lukose, R.~M. (2008).
\newblock Budget constrained bidding in keyword auctions and online knapsack
  problems.
\newblock In {\em WINE}.

\end{thebibliography}

\appendix

\section{Simulation Detail}\label{appendix:simulation_detail}

\begin{figure}[H]
\centering
\includegraphics[scale=0.5]{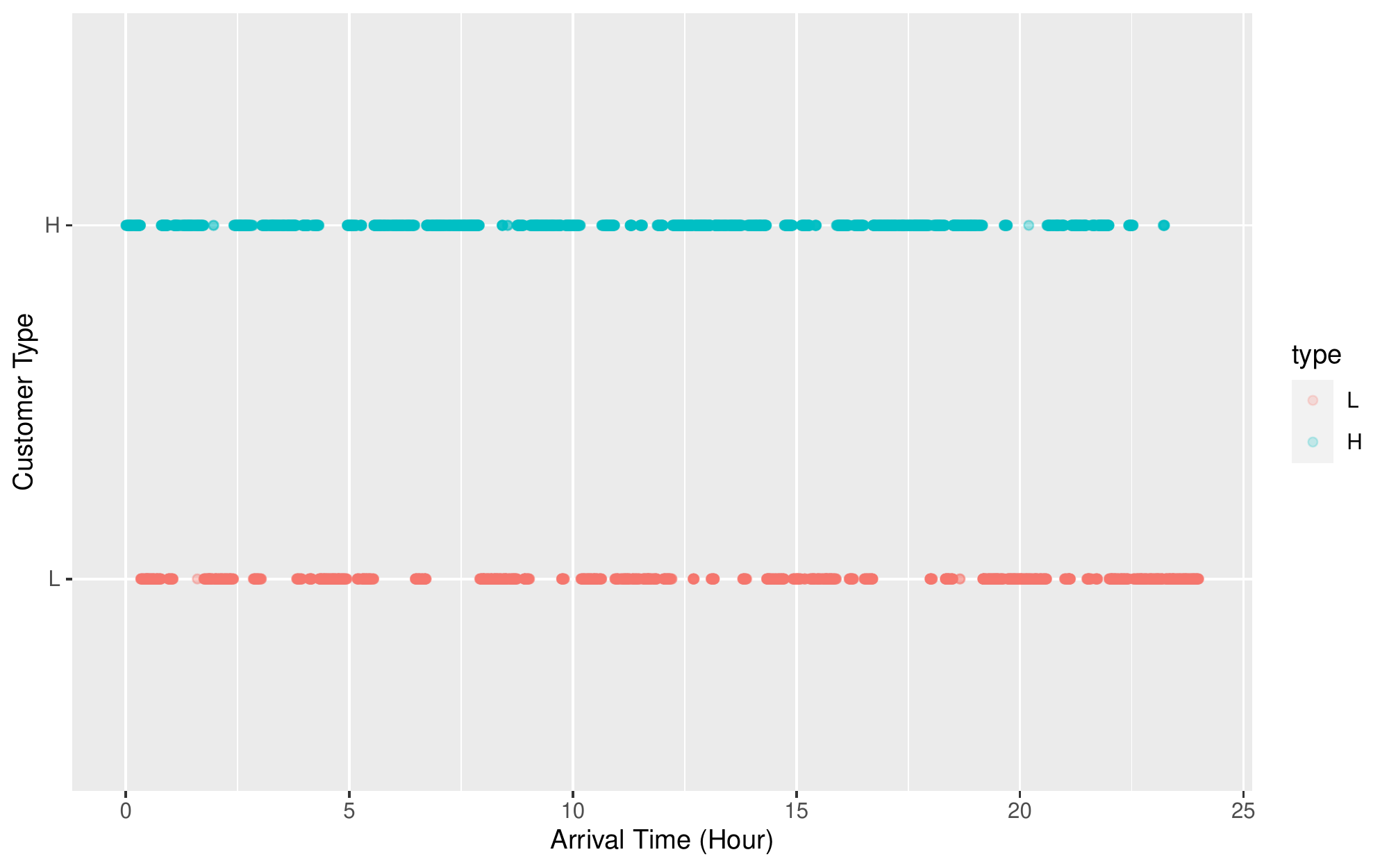}
\caption{Customer Type Vs Arrival Time}\label{fig:consumer_type}
\end{figure}

Figure \ref{fig:consumer_type} shows the arrival of consumers. Because of auto-correlation, similar consumers tend to arrive around similar time. 

\begin{figure}[H]
\centering
\includegraphics[scale=0.5]{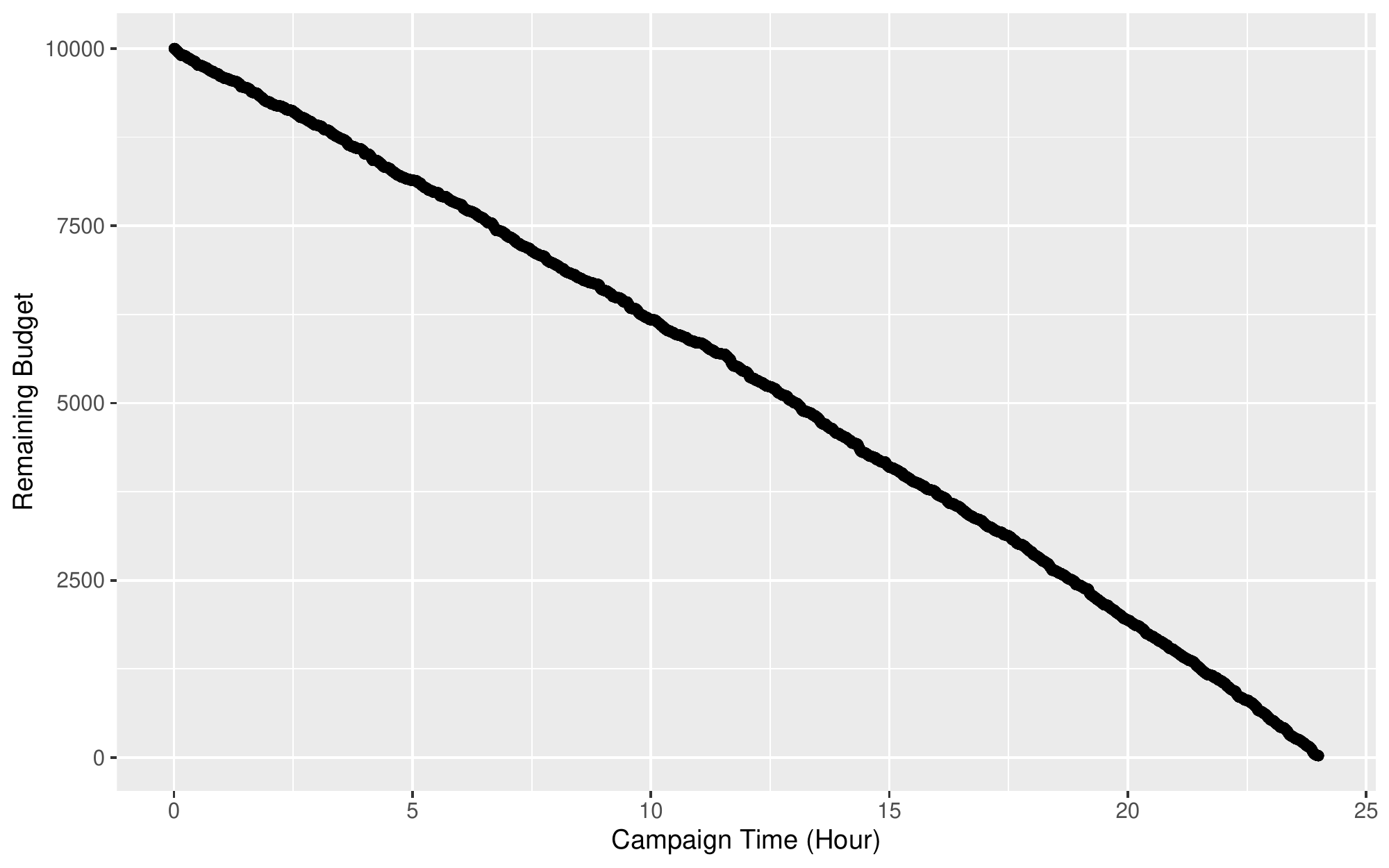}
\caption{Remaining Budget Vs Arrival Time}\label{fig:budget_evolution}
\end{figure}

Figure \ref{fig:budget_evolution} shows how budget evolves over time. Because of the throttling algorithm, the budget did not exhaust until the end of the campaign. 

\begin{figure}[H]
\centering
\includegraphics[scale=0.5]{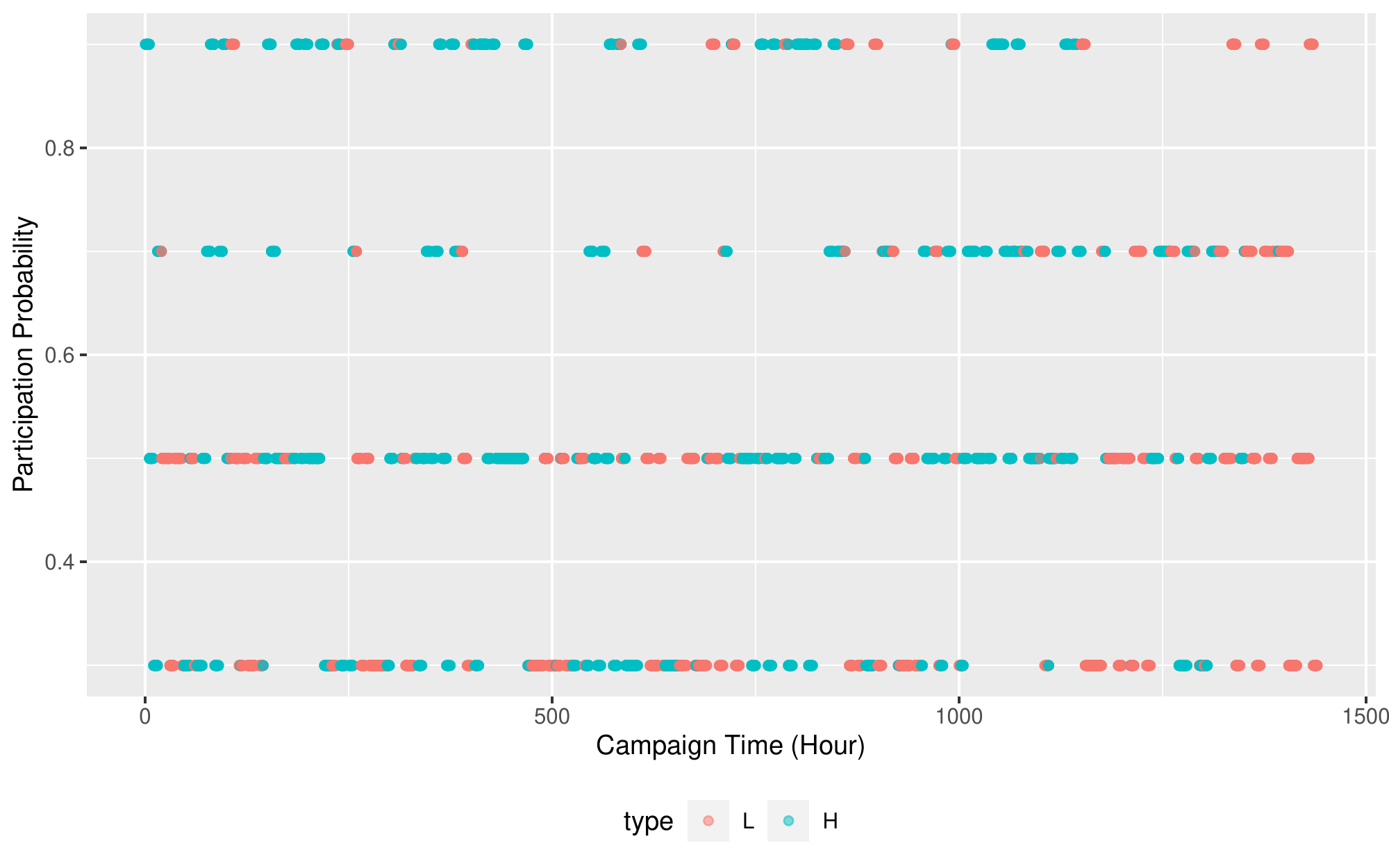}
\caption{Remaining Budget Vs Arrival Time}\label{fig:prob_evolution}
\end{figure}

Figure \ref{fig:prob_evolution} shows how the participation probability changes over time due to the change in expenditure rate in the last interval. 

\begin{figure}[H]
\centering
\includegraphics[scale=0.5]{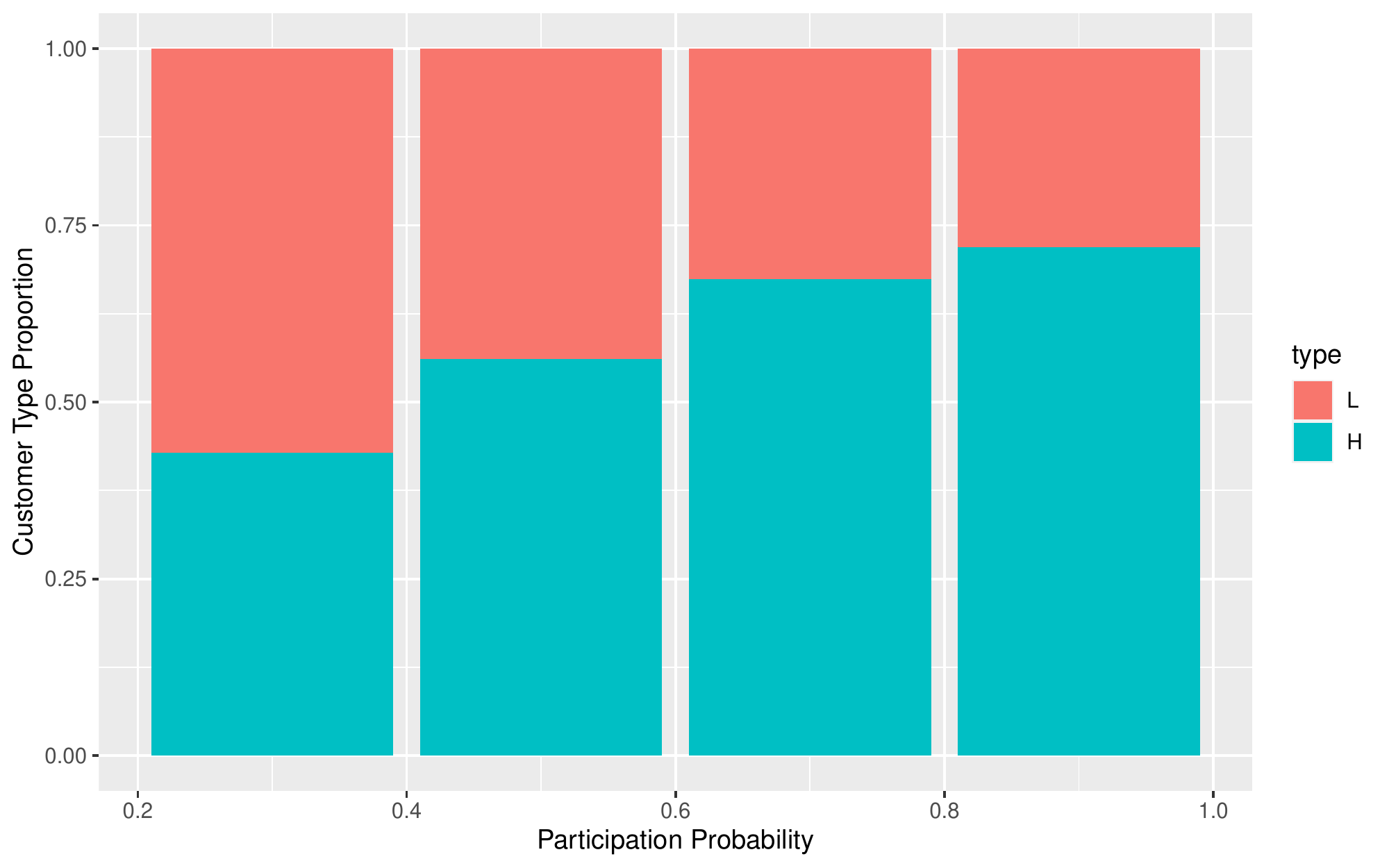}
\caption{Remaining Budget Vs Arrival Time}\label{fig:type_proportion}
\end{figure}

Such change is mainly driven by the type of customers, as illustrated in Figure \ref{fig:type_proportion}

%
%

\section{Inference Assuming Fixed Participation Probability}\label{sec:fixed_prob}

The inference would be easy if the participation probability $p_i$ is a priori fixed, and the realized participation $Z_i$ is the only source of variation. 
Under the assumption that participation probabilities are fixed, our statistical inference problem reduces to a stratified experiment with partial compliance, in which strata are defined by participation probabilities. 

\begin{prop} \label{prop: asymptotic normality}
{\bf Suppose the participation probabilities are a priori fixed. }
Let $n = \sum_{p \in \cP} n_p$ be the total sample size. 
As $n$ and $n_p$s go to infinity at the same rate, the estimator $\hat{\tau}_{\LATE}$ is asymptotically normal: 
\begin{align*}
\sqrt{n}\left(\hat{\tau}_{\LATE} - \tau_{\LATE}\right)
\overset{d}{\rightarrow} 
\Normal\left(0, \Omega\right),
\end{align*}
in which the asymptotic variance is 
\begin{align*}
\Omega & 
= \sum_{p \in \cP}w_p^2 \cdot \frac{\Sigma_{Y, p} + \Sigma_{D, p} \cdot \tau_{\LATE}^2 - 2 \Sigma_{YD, p} \cdot \tau_{\LATE}}{\ITT_{D, p}^2},
\end{align*}
where $\Sigma_{Y, p} = \lim n \cdot \var\left(\ITThat_{Y, p}\right)$, $\Sigma_{D, p} = \lim n \cdot \var\left(\ITThat_{D, p}\right)$, $\Sigma_{YD, p}= \lim n \cdot \cov\left(\ITThat_{Y, p}, \ITThat_{D, p}\right)$.
\end{prop}
\begin{proof}
We outline the proof in the following . 
Under proper regularity conditions for a central limit theorem, we have 
\begin{align*}
\sqrt{n}\left(\ITThat_{Y, p} - \ITT_{Y, p}, \ITThat_{D, p} - \ITT_{D, p}, p \in \cP\right) 
\overset{d}{\rightarrow}
\Normal\left(0, \operatorname{diag}(\Sigma_p, p\in \cP)\right),
\end{align*}
in which 
\begin{align*}
\Sigma_p & = \lim n \cdot 
\begin{pmatrix}
\var\left(\ITThat_{Y, p}\right) & \cov\left(\ITThat_{Y, p}, \ITThat_{D, p}\right)\\
\cov\left(\ITThat_{Y, p}, \ITThat_{D, p}\right) & \var\left(\ITThat_{D, p}\right)
\end{pmatrix} = \begin{pmatrix}
\Sigma_{Y, p} & \Sigma_{YD, p}\\
\Sigma_{YD, p} & \Sigma_{D, p}
\end{pmatrix}.
\end{align*}
Apply delta method to the following expression
\begin{equation*}
\hat{\tau}_{\LATE} 
= g\left(\ITThat_{Y, p}, \ITThat_{D, p}, p \in \cP\right)
= \frac{\sum_{p \in \cP} n_p \cdot \ITThat_{Y, p}}{\sum_{p \in \cP} n_p \cdot \ITThat_{D, p}}.
\end{equation*}
We get
\begin{align*}
\sqrt{n}\left(\hat{\tau}_{\LATE} - \tau_{\LATE}\right)
\overset{d}{\rightarrow} 
\Normal\left(0, \Omega\right),
\end{align*}
in which the asymptotic variance is 
\[
\Omega = \nabla g\operatorname{diag}(\Sigma_p, p\in \cP)\nabla g^\top
\]
where the gradient is 
\begin{align*}
\nabla g\left(\ITT_{Y, p}, \ITT_{D, p}, p \in \cP\right)
& = \left(\frac{n_p}{\sum_{q\in \cP} n_q  \cdot \ITT_{D, q} }, -\frac{\sum_{q\in \cP}  n_q  \cdot \ITT_{Y, q}}{\left(\sum_{q\in \cP} n_q  \cdot \ITT_{D, q}\right)^2}\cdot n_p , p \in \cP\right)\\
& = \left(\frac{w_p}{\ITT_{D, p}}, - \frac{w_p}{\ITT_{D, p}} \tau_{\LATE}, p \in \cP\right).
\end{align*}
More explicitly, 
\begin{align*}
\Omega & = \lim n \cdot \sum_{p \in \cP}w_p^2 \cdot \var\left(\frac{\ITThat_{Y, p} - \ITThat_{D, p}\cdot \tau_{\LATE}}{\ITT_{D, p}}\right)\\
& = \sum_{p \in \cP}w_p^2 \cdot \frac{\Sigma_{Y, p} + \Sigma_{D, p} \cdot \tau_{\LATE}^2 - 2 \Sigma_{YD, p} \cdot \tau_{\LATE}}{\ITT_{D, p}^2}.
\end{align*}
\end{proof}
To draw inference based on this asymptotic normality result, we need to estimate the asymptotic variance. 
Define the following estimator
\begin{align}
\widehat{\Omega}
& = \sum_{p \in \cP}\hat{w}_p^2 \cdot \frac{\widehat{\Sigma}_{Y, p} + \widehat{\Sigma}_{D, p} \cdot \hat{\tau}_{\LATE}^2 - 2 \widehat{\Sigma}_{YD, p} \cdot \hat{\tau}_{\LATE}}{\widehat{\ITT}_{D, p}^2}, \label{eq: variance estimator}
\end{align}
where $\hat{\Sigma}_{t, Y}$, $\hat{\Sigma}_{t, D}$, and $\hat{\Sigma}_{t, YD}$ are defined as 
\begin{align*}
\widehat{\Sigma}_{Y, p} & = n\cdot \left[\frac{1}{n_{p, 1}}\widehat{\var}_p(Y_i(D_i(1))) + \frac{1}{n_{p, 0}}\widehat{\var}_p(Y_i(D_i(0)))\right]\\ 
\widehat{\Sigma}_{D, p} & = n\cdot \frac{1}{n_{p, 1}}\widehat{\var}_p(D_i(1))\\ 
\widehat{\Sigma}_{YD, p} & = n\cdot \frac{1}{n_{p, 1}}\widehat{\cov}_p\left(Y_i(D_i(1)), D_i(1)\right).
\end{align*}
The terms on the right-hand side of the equations are
\begin{align*}
\widehat{\var}_p(Y_i(D_i(1))) & = \frac{1}{n_{p, 1} - 1} \sum_{i\in \cI_{p, 1}} 
\left(Y_i - \frac{1}{n_{p, 1}} \sum_{i\in \cI_{p, 1}} Y_i\right)^2\\
\widehat{\var}_p(Y_i(D_i(0))) & = \frac{1}{n_{p, 0} - 1} \sum_{i\in \cI_{p, 0}} 
\left(Y_i - \frac{1}{n_{p, 0}} \sum_{i\in \cI_{p, 0}} Y_i\right)^2\\
\widehat{\var}_p(D_i(1)) & = \frac{1}{n_{p, 1} - 1} \sum_{i\in \cI_{p,1}} 
\left(D_i - \frac{1}{n_{p, 0}} \sum_{i\in \cI_{p, 1}} D_i\right)^2\\
\widehat{\cov}_p\left(Y_i(D_i(1)), D_i(1)\right) & = \frac{1}{n_{p, 1} - 1} \sum_{i\in \cI_{p, 1}} 
\left(Y_i - \frac{1}{n_{p, 1}} \sum_{i\in \cI_{p, 1}} Y_i\right) \cdot 
\left(D_i -  \frac{1}{n_{p, 1}} \sum_{i\in \cI_{p, 1}} D_i\right).
\end{align*}
The variance estimator presented in the main text and reported in the simulation is 
\[
\widehat{\var}\left(\hat{\tau}_{\LATE}\right) = \frac{\widehat{\Omega}}{n}.
\]
The $95\%$ confidence interval used in the simulation is the following
\begin{equation}
\left[\hat{\tau}_{\LATE} - 1.96\cdot \sqrt{\widehat{\var}\left(\hat{\tau}_{\LATE}\right)}, \hat{\tau}_{\LATE} + 1.96\cdot 
\sqrt{\widehat{\var}\left(\hat{\tau}_{\LATE}\right)}
\right].  \label{eq: analytical CI}
\end{equation}



The assumption of non-stochastic participation probabilities $p$ is plausible when the algorithm dictates the schedule of the participation probability before the campaign starts: for example, the participation probability is scheduled to be $90\%$ before $3pm$ and $70\%$ after $3pm$, regardless of the remaining budget. 

Though the assumption of non-stochastic participation probability does not hold for throttling algorithms that adaptively adjust participation probabilities, the inference method will still work well when unpredictable variation in participation probabilities is only moderate. 
In particular, this inference method often works well for large advertising campaigns, of which participation probabilities follow a predictable temporal pattern.

In our simulation exercise, because the sample size is large and our algorithm uses average expenditure as the main input, the additional design-based uncertainty induced by the variation of the participation probability is small. Because some of the variance estimator for $\widehat{\ITT}$s are conservative when there is no variation in the participation probability, the confidence interval is conservative when the variation in participation probability is small.

\end{document}